\documentclass[%
 reprint,
superscriptaddress,
bibnotes,
amsmath,amssymb,
aps,
pra,
]{revtex4-1}

\usepackage{graphicx,import}
\usepackage{dcolumn}
\usepackage{braket}
\usepackage{textcomp, gensymb}
\usepackage{bm}
\usepackage{amsmath}
\usepackage{float}
\usepackage{mathtools}
\usepackage{color}
\usepackage[dvipsnames,table,xcdraw]{xcolor}
\usepackage[utf8]{inputenc}
\usepackage{csquotes}
\usepackage{thmtools}

\usepackage{dsfont}
\usepackage{bbm}

\usepackage{hyperref}

\usepackage[normalem]{ulem}

\usepackage{outline}
\usepackage{physics}
\usepackage{multirow}
\usepackage{siunitx}
\usepackage{tikz}
\usetikzlibrary{shapes.geometric, arrows,positioning}
\usepackage{qcircuit}
\usepackage{lipsum}
\usepackage{chemfig}
\usepackage{silence}
\usepackage{algorithm}
\usepackage[noend]{algpseudocode}
\DeclareMathOperator*{\argmax}{argmax}
\DeclareMathOperator*{\argmin}{argmin}

\makeatletter
\def\p@subsection{}
\makeatother
\makeatletter
\def\p@subsubsection{}
\makeatother

\allowdisplaybreaks

\newcommand{\pmax}{P}

\newcommand{\pool}{\mathcal{P}}
\newcommand{\bound}{\mathcal{E}}
\newcommand{\bigO}{\mathcal{O}}
\newcommand{\fluc}{\delta E}
\renewcommand{\grad}{\mathcal{G}}

\newcommand{\Vmax}{V_{\text{max}}}
\newcommand{\VGW}{V_{\text{GW}}}
\newcommand{\VS}{V_{\text{s}}}
\newcommand{\VD}{V_{\text{d}}}

\newcommand{\AGW}{\alpha_{\text{GW}}}
\newcommand{\AS}{\alpha_{\text{s}}}
\newcommand{\AD}{\alpha_{\text{d}}}

\newcommand{\pgate}{p_{\text{gate}}}

\begin{document}

\title{Dynamic-ADAPT-QAOA: An algorithm with shallow and  
noise-resilient circuits}
\author{Nikola Yanakiev}
\affiliation{Hitachi Cambridge Laboratory, J. J. Thomson Ave., Cambridge, CB3 
0HE, United Kingdom}
\affiliation{Cavendish Laboratory, Department of Physics, University of 
Cambridge, Cambridge, CB3 0HE, United Kingdom}

\author{Normann Mertig}
\affiliation{Hitachi Cambridge Laboratory, J. J. Thomson Ave., Cambridge, CB3 
0HE, United Kingdom}

\author{Christopher K. Long}
\affiliation{Hitachi Cambridge Laboratory, J. J. Thomson Ave., Cambridge, CB3 
0HE, United Kingdom}
\affiliation{Cavendish Laboratory, Department of Physics, University of 
Cambridge, Cambridge, CB3 0HE, United Kingdom}

\author{David R. M. Arvidsson-Shukur}
\affiliation{Hitachi Cambridge Laboratory, J. J. Thomson Ave., Cambridge, CB3 
0HE, United Kingdom}

\date{\today}

\begin{abstract}
The quantum approximate optimization algorithm (QAOA) is an appealing proposal 
to solve NP problems on noisy intermediate-scale quantum (NISQ) hardware.
Making NISQ implementations of the QAOA resilient to noise requires short ansatz 
circuits with as few CNOT gates as possible.
Here, we present Dynamic-ADAPT-QAOA.
Our algorithm significantly reduces the circuit depth 
and the CNOT count of standard ADAPT-QAOA, a leading 
proposal for near-term implementations of the QAOA.
Throughout our algorithm, the decision to apply CNOT-intensive operations is made dynamically, based on algorithmic benefits.
Using density-matrix simulations, we benchmark the noise resilience of ADAPT-QAOA and Dynamic-ADAPT-QAOA. We compute the gate-error 
probability $\pgate^\star$ below which these algorithms
provide, on average, more accurate solutions than the classical, polynomial-time approximation algorithm by Goemans and Williamson.
For small systems with $6-10$ qubits, we show that $\pgate^\star>10^{-3}$ 
for Dynamic-ADAPT-QAOA. Compared to standard ADAPT-QAOA, this constitutes 
an order-of-magnitude improvement in
noise resilience.
This improvement should make Dynamic-ADAPT-QAOA viable for implementations on 
superconducting NISQ hardware, even in the absence of error mitigation.
\end{abstract}

\maketitle

\section{Introduction}
\label{sec:introduction}

NP problems are ubiquitous in computer science, occurring frequently in 
combinatorial optimization, and machine learning \cite{np_bible,UBQP}.
Finding their solutions is computationally hard.
One strategy to solve NP problems, relies on the Ising model 
\cite{ising_formulations, spin_glass_np,ising_model_translation}.
An NP problem is encoded in the real and symmetric matrix $W_{ij}$.
The (approximate) solution is then found by approximating the ground state 
energy $E_0$ of an Ising Hamiltonian
\begin{align}
    H = \frac{1}{4}\sum_{i,j=1}^{N} W_{ij}Z_i Z_j,
    \label{eq:cost_ham}
\end{align}
where $Z_i$ denotes the Pauli-$z$ operator acting on qubit $i=1,\dots,N$.
Approximate solutions are usually found using heuristics 
\cite{simulated_annealing_1, simulated_annealing_2,simulated_annealing_3, 
momentum_annealing} or adiabatic quantum computers 
\cite{quantum_annealing,adiabatic_quantum_computation,manufactured_spins, 
quantum_adiabatic_evolution}.
The quality of these solutions can be assessed using the Goemans and Williamson 
(GW) algorithm \cite{goemans_williamson}, which, in the worst case, provides approximate solutions 
within $87.8\ldots \%$ of the true ground-state energy in polynomial time 
(using an alternative representation of the NP problem).

Recent works \cite{qaoa,oh2019solving} have proposed solving NP problems on gate-based 
quantum computers, using the quantum approximate optimization algorithm (QAOA).
The QAOA identifies approximate solutions to NP problems by creating upper bounds to 
the ground-state energy $E_0$ of $H$ via the Rayleigh-Ritz variational 
principle:
\begin{align}
 E_0 \le E(\vec{\beta},\vec{\gamma}) = 
\langle \Psi(\vec{\beta},\vec{\gamma}) | H 
|\Psi(\vec{\beta},\vec{\gamma})\rangle.
\end{align}
The classically-hard-to-represent trial state is prepared on a quantum computer 
by evolving an initial state $|\Psi_0\rangle$:
\begin{equation}
    |\Psi(\vec{\beta},\vec{\gamma})\rangle = 
    U_{\pmax}(\vec{\beta},\vec{\gamma})|\Psi_0\rangle,
\label{eq:qaoa_trial_state}
\end{equation}
using a parametrized ansatz circuit
\begin{equation}
    U_{\pmax}(\vec{\beta},\vec{\gamma}) = 
 \prod_{p=1}^{\pmax}\left[e^{-i\beta_p A_p}e^{-i\gamma_p H}\right].
\label{eq:qaoa_unitary}
\end{equation}
The QAOA then optimizes the parameters to minimize the energy expectation value 
$E(\vec{\beta},\vec{\gamma})$.

In the original proposal of QAOA \cite{qaoa}, the form of the ansatz circuit 
[Eq.~\eqref{eq:qaoa_unitary}] is inspired by a Trotterized form of 
the adiabatic theorem \cite{Born1928}. 
By setting the mixer Hamiltonian to $A_p=\prod_{i=1}^{N} X_i$ for all 
$p$, and the initial state to $\ket{\Psi_0}=\ket{+}...\ket{+}$, the QAOA finds 
the ground state exactly as the number of Trotter steps tends to infinity 
$(\pmax\rightarrow\infty)$.
Unfortunately, large values of $\pmax$ lead to intractably deep ansatz 
circuits.
In the presence of noise, the need for deep circuits precludes the 
implementation of the QAOA on existing quantum hardware \cite{noisy_limitations,noisy_limitations_2}.

To reduce the intractably deep quantum circuits, ADAPT-QAOA \cite{adapt_qaoa} was
developed.
The algorithm improves the ansatz circuit in $\pmax$ iterations.
Further, it allows the mixer Hamiltonian $A_p$ to vary in each iteration $p$, 
by choosing it from a mixer pool $\pool$.
In noiseless numerical simulations, ADAPT-QAOA has been demonstrated to 
generate shallower circuits than the QAOA.
Despite these improvements, ADAPT-QAOA lies outside the scope of current 
hardware.
Moreover, the resilience of ADAPT-QAOA to noise has never been 
quantified.

In this paper, we benchmark ADAPT-QAOA in the presence of noise.
Using density-matrix simulations, we compute the gate-error probability 
$\pgate^\star$ below which the quantum algorithm  outputs, on average,
better approximate solutions than the classical GW algorithm.
For small systems of $6-10$ qubits, we find that ADAPT-QAOA 
requires $\pgate^\star$ comparable to or smaller than the gate-error 
probabilities available on current 
hardware.
To reduce the hardware requirements of ADAPT-QAOA further, we develop 
Dynamic-ADAPT-QAOA.
This  algorithm  removes redundant components from the ansatz circuits.
For the problems we study, Dynamic-ADAPT-QAOA reduces the circuit depths 
significantly. For instance, in noiseless simulations of 
6-qubit systems, Dynamic-ADAPT-QAOA  achieves a better average performance than the GW algorithm with  
approximately $80\%$ fewer CNOT gates than the original ADAPT-QAOA.
This reduction in CNOT gates leads to improved noise resilience, with 
$\pgate^\star$ being approximately an order of magnitude better than that 
of the original ADAPT-QAOA.
Dynamic-ADAPT-QAOA may thus be implementable on current superconducting 
hardware, even in the absence of error mitigation.

\section{Dynamic-ADAPT-QAOA}
\label{sec:algos}

In this section, we introduce Dynamic-ADAPT-QAOA.
Our presentation strategy is to first review the standard ADAPT-QAOA template.
Subsequently, we describe its improvement via Dynamic-ADAPT-QAOA.

\subsection{ADAPT-QAOA}

\begin{figure}
\includegraphics[width=\columnwidth]{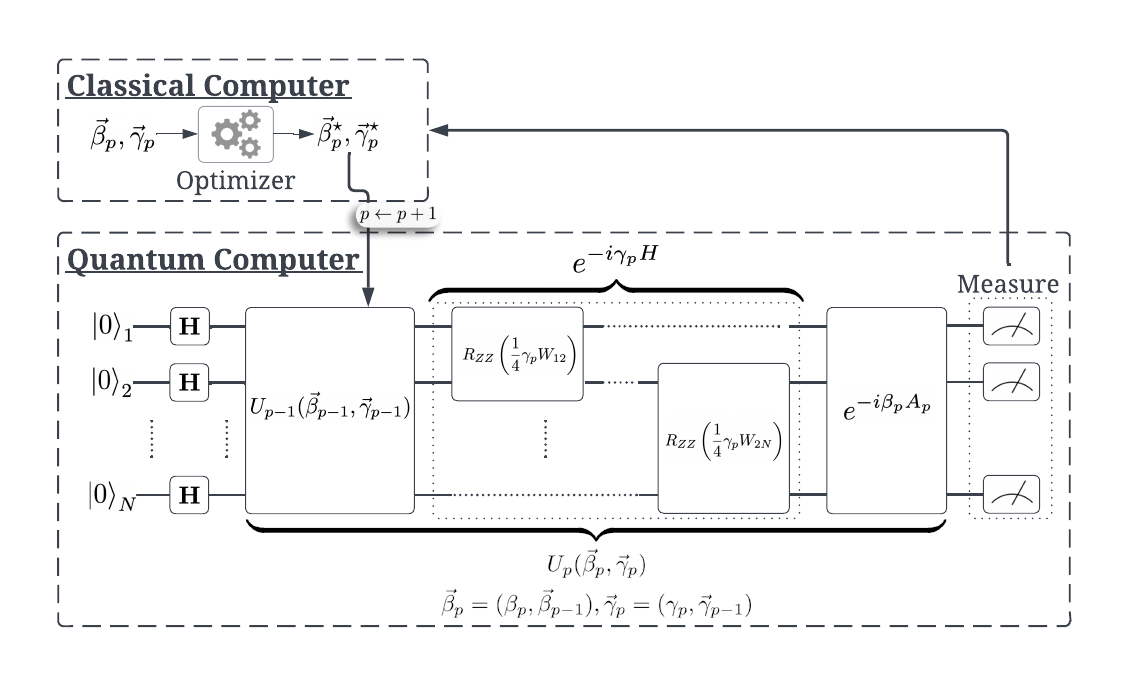}
    \caption{The $p$th iteration of ADAPT-QAOA: After initialization, the 
ansatz circuit from the previous iteration $U_{p-1}$ is augmented by appending 
unitary evolutions generated by $H$ and $A_p$. The optimal circuit parameters 
$\vec{\beta}_p^{\star}, \vec{\gamma}_p^{\star}$ are identified by minimizing 
the measured energy expectation.}
\label{fig:ADAPTQAOA}
\end{figure}

As depicted in Fig.~\ref{fig:ADAPTQAOA}, ADAPT-QAOA grows the ansatz circuit in 
$\pmax$ steps.
In each step $p$, unitary evolutions generated by $H$ and $A_p$ are appended to 
the circuit from the previous step:
\begin{align}
    \label{eq:CircuitUpdate}
 U_{p}(\vec{\beta}_{p}, \vec{\gamma}_{p}) = 
 e^{-i\beta_p A_p}e^{-i\gamma_p H} U_{p-1}(\vec{\beta}_{p-1}, 
\vec{\gamma}_{p-1}).
\end{align}
The process starts from $U_0=\text{\normalfont id}$.
Concurrently, the real parameter vectors are updated as
\begin{align}
 \label{eq:ParameterVector}
 \vec{\beta}_p = (\beta_p, \vec{\beta}_{p-1})
 \quad\text{and}\quad
 \vec{\gamma}_p = (\gamma_p, \vec{\gamma}_{p-1}),
\end{align}
starting from empty vectors $\vec{\beta}_0=()$ and $\vec{\gamma}_0=()$.
In each step, an optimal mixer Hamiltonian $A_p$ is picked from a pool $\pool$ such that the energy gradient is maximized  (see below).
The circuit parameters are then optimized:
\begin{align}
 \vec{\beta}_p^{\star},\vec{\gamma}_p^{\star} =
     \argmin_{\vec{\beta}_p,\vec{\gamma}_p}
          \left[E_p(\vec{\beta}_p,\vec{\gamma}_p)\right],
\end{align}
to minimize the energy expectation value
\begin{align}
    \label{eq:EnergyExpectation}
 E_p(\vec{\beta}_p,\vec{\gamma}_p) = 
 \bra{\Psi_0} U_{p}^{\dagger}(\vec{\beta}_{p},\vec{\gamma}_{p}) H
U_{p}(\vec{\beta}_{p},\vec{\gamma}_{p})\ket{\Psi_0}.
\end{align}
This yields an upper bound
$\bound_p = E_p(\vec{\beta}_p^{\star},\vec{\gamma}_p^{\star})$
on the ground-state energy $E_0$, and an optimal trial state
$|\Psi_{p}^{\star}\rangle \equiv U_{p}(\vec{\beta}_{p}^{\star}, 
\vec{\gamma}_{p}^{\star})|\Psi_0\rangle$.
Iterating this process, provides a hierarchy of bounds $\bound_0 > \bound_1 > 
\cdots > \bound_p > \cdots \ge E_0$.
The algorithm terminates, when $p=\pmax$ or if $|\bound_{p-1}-\bound_p|$ falls 
below a pre-defined threshold $\varepsilon$.

To accelerate convergence, ADAPT-QAOA picks the mixer Hamiltonian which 
maximizes the energy gradient.
To evaluate this gradient, the optimal trial state is augmented by appending a 
cost and a mixer unitary:
\begin{align}
    \label{eq:state_variation}
 \Ket{\Psi_p(\beta_p,\gamma_p; A)} 
 = e^{-i\beta_p A}e^{-i\gamma_p H} \Ket{\Psi_{p-1}^{\star}}.
\end{align}
The energy variation due to the added parameters
\begin{align}
    \label{eq:two_dim_func}
  \fluc_p(\beta_p,\gamma_p; A) = 
\Bra{\Psi_p(\beta_p,\gamma_p; A)} H 
\Ket{\Psi_p(\beta_p,\gamma_p; A)},
\end{align}
enables the definition of a corresponding energy gradient:
\begin{equation}
    \grad_p(\gamma_p; A)\equiv 
    \left.
    \frac{\partial}{\partial{\beta_p}}  \fluc_p(\beta_p,\gamma_p; A)\right|_{\beta_p=0}.
     \label{eq:gradient_exp}
 \end{equation}
Evaluating this gradient for each $A\in\pool$ allows for selecting the optimal 
mixer:
\begin{align}
    \label{eq:optimal_mixer}
 A_p = \argmax_{{A}\in\mathcal{P}}\left[|\grad_p(\gamma_p; {A})|\right].
\end{align}
Throughout this work, we use the same mixer pool as in ADAPT-QAOA \cite{adapt_qaoa}, 
comprising of QAOA mixers as well as Pauli strings of length one and two:
\begin{align}
 \pool
    &= \left\{\sum_{i=1}^N X_i, \sum_{i=1}^N Y_i\right\}
    \cup \left\{X_i, Y_i \,|\, i=1,...,N \right\} \\ \nonumber
    &\cup \left\{\sigma_i \sigma'_j \,|\, \sigma,\sigma'\in\{X,Y,Z\}\land i,j=1,...,N\land i\neq j 
           \right\}.
\end{align}

\subsection{Dynamic-ADAPT-QAOA}
\label{sec:dADAPT}

\textit{Motivation:---}Our motivation for developing Dynamic-ADAPT-QAOA 
comes from two observations.
First, In each step $p$, the quantum circuit representing the cost unitary 
$e^{-i\gamma_p H}$ requires $\bigO(N^2)$ CNOT gates (see App.~\ref{app:CNOT_count}).
On the other hand, the quantum circuit representing the mixer unitary $e^{-i\beta_p 
A_p}$ requires only $\bigO(1)$ CNOT gates \cite{mixer_unitary_implementations}.
As CNOT gates induce noise, minimizing the number of  cost unitaries in the ansatz circuit could be valuable \cite{layering_noise}.
Second, in standard ADAPT-QAOA, the vector of optimal parameters 
$\vec{\gamma}_{p}^{\star}$ tends to be sparse, with many parameters taking 
values close to zero (see Sec.~\ref{sec:vanish_cost_param}).
As cost unitaries $e^{-i\gamma_p H}$ with $\gamma_p\approx0$ hardly affect the final quantum circuit, it could be advantageous to exclude them altogether.

\textit{Idea:---}In general, the energy expectation value in Eq.~\eqref{eq:EnergyExpectation} is 
a nontrivial function of the circuit parameters.
Hence, it is not obvious how to predict which entries in 
$\vec{\gamma}_p^\star$  would take optimal values close to zero.
Yet, in ADAPT-QAOA, optimal circuit parameters of the $p$th iteration are usually 
well approximated by the circuit parameters of the previous iteration:
\begin{align}
 \vec{\beta}_p^\star  \approx (\beta_p^\star, \vec{\beta}_{p-1}^\star)
 \quad\text{and}\quad
 \vec{\gamma}_p^\star \approx (\gamma_p^\star, \vec{\gamma}_{p-1}^\star).
\end{align}
Thus, we can estimate the optimal circuit parameters $\beta_p^\star, 
\gamma_p^\star$ of the $p$th iteration, by studying the minima of 
\begin{equation}
\label{eq:two_dim_func_opt_mixer}
  \fluc_p(\beta_p,\gamma_p) \equiv \fluc_p(\beta_p,\gamma_p;{A}_p).
\end{equation}
As explained in App.~\ref{Appendix:EnergyMinimaAnalysis}, for Pauli-string 
mixers $A_p$, we can identify whether $\fluc_p(\beta_p,\gamma_p)$ has minima 
near $\gamma_p^\star=0$.
To this end, we split the cost Hamiltonian into two parts $H = H_{-} + H_{+}$, 
such that $H_{-}$ commutes and $H_{+}$ anti-commutes with $A_p$.
This enables the evaluation of three additional expectation values:
\begin{subequations}
  \label{eq:param_space_quantities}
  \begin{align}
  B_p &= \Bra{\Psi_{p-1}^\star}{i{A}_p{H}_{+}}\Ket{\Psi_{p-1}^\star}
         \equiv\grad_p(0;A_p),\\
  C_p &= \Bra{\Psi_{p-1}^\star}{{A}_p{H}_{+}^2}\Ket{\Psi_{p-1}^\star},\\
  D_p &= \Bra{\Psi_{p-1}^\star}{i{A}_p{H}_{+}^3}\Ket{\Psi_{p-1}^\star}.
  \end{align}
\end{subequations}
As shown in App.~\ref{Appendix:EnergyMinimaAnalysis}, $\fluc_p(\beta_p, 
\gamma_p)$ has a local minimum at $\gamma_p^\star=0$ if
\begin{align}
 \label{eq:Minima}
 C_p=0 \quad\text{and}\quad B_p D_p > 0.
\end{align}

\textit{Algorithm:---}Dynamic-ADAPT-QAOA excludes the cost unitary of the 
$p$th iteration if $A_p$ is a Pauli-string and Condition~\eqref{eq:Minima} 
holds.
Otherwise, the algorithm follows the standard mixer-selection procedure of 
ADAPT-QAOA.
That is, the gradients for all $A\in\pool$ are re-evaluated at some given offset
$\gamma_p=\pm\tilde{\gamma}$, and the optimal mixer is determined:
\begin{align}
    A_p = 
\argmax_{A\in\pool}
\left[\max(|\grad_p(+ \tilde{\gamma};A)|, |\grad_p(-\tilde{\gamma};A)|)\right].
\end{align}
After determining ${A}_p$, the ansatz circuit and parameter vectors are 
grown as described in Eqs.~\eqref{eq:CircuitUpdate} and~\eqref{eq:ParameterVector}.
Pseudocode summarizing Dynamic-ADAPT-QAOA is given in 
Algorithm~\ref{alg:dynamic_adapt_qaoa}.

\textit{Remarks:---}In App.~\ref{app:algo_subtleties}, we discuss two alterations of Dynamic-ADAPT-QAOA. In the first alteration, all cost unitaries are, \textit{a priori}, removed from the ansatz circuit. In the second alteration, the algorithm does not re-evaluate the optimal mixer $A_p$ at $\gamma_p =\pm\tilde{\gamma}$ if condition (17) fails.  As shown in App.~\ref{app:algo_subtleties}, both of these alterations worsen the algorithmic performance.

Common worries regarding variational quantum algorithms concern barren plateaus (vanishing gradients)  and the presence of bad local minima \cite{McClean18,Wiersema20, Kiani20, Cerezo21, Marrero21, You21, Anschuetz22, Kiani22}.  A promising way to mitigate these issues is to reduce the circuit depths \cite{Kiani22, Deshpande22}, which is precisely what our algorithm does.  Moreover,  since the gates of adaptive variational quantum algorithms are  tailored to the optimization problem itself,  there are indications that these algorithms avoid such issues better than other variational quantum algorithms \cite{Grimsley19, Yordanov21, Kiani22, dalton2022variational, Grimsley23}.    In the instances studied below, Dynamic-ADAPT-QAOA efficiently implements the variational optimization.

\begin{algorithm}[H]
    \caption{Dynamic-ADAPT-QAOA}\label{alg:dynamic_adapt_qaoa}
    \begin{algorithmic}[]
        \State Init pool $\mathcal{P}$;
               state $\ket{\Psi_0} \gets \ket{+}\dots\ket{+}$; 
               unitary ${U}_0\gets{I}$.
        \State Init accuracies $\varepsilon, \delta_1, \delta_2$;
               and offset $\tilde{\gamma}$.
        \State Init optimal params $\vec{\beta}_{0}^\star\gets(), 
                                   \vec{\gamma}_{0}^\star\gets()$;
               Init $p \gets 1$.
        \While{not converged}
        \State Prepare $\ket{\Psi_{p-1}^\star} 
                        \gets {U}_{p-1}(\vec{\beta}_{p-1}^\star,
                              \vec{\gamma}_{p-1}^\star)\ket{\Psi_0}$
        \State Evaluate gradients $\grad_p({A})\gets
           \Braket{\Psi_{p-1}^\star|\left[i{A},{H}\right]|\Psi_{p-1}^\star}$
        \State Find optimal mixer: 
           ${A}_p\gets\argmax_{{A}\in\mathcal{P}}\left[|\grad_p(0;A)|\right]$
        \State Evaluate $B_p$, $C_p$, $D_p$ in 
                                          Eq.(\ref{eq:param_space_quantities})
        \If{$|C_p|\leq\delta_1$ \textbf{and} $B_p\cdot D_p>\delta_2$}
        \State Update $\vec{\gamma}_p\gets\vec{\gamma}_{p-1}$;
               $\vec{\beta}_p\gets(\beta_p,\vec{\beta}_{p-1})$
        \State Append ${U}_p(\vec{\beta}_p,\vec{\gamma}_p) \gets 
             e^{-i\beta_p{A}_p}{U}_{p-1}(\vec{\beta}_{p-1},\vec{\gamma}_{p-1})$
        \Else
        \State Prepare $|\tilde{\Psi}_p^\pm\rangle \gets e^{\mp 
                        i\tilde{\gamma}{H}}|\Psi_{p-1}^\star\rangle$
        \State Measure gradient
               $\grad_p(\pm\tilde{\gamma}, A)\gets
               \langle\tilde{\Psi}_{p}^\pm|\left[i{A},{H}\right ] |
               \tilde {\Psi}_{p}^\pm\rangle$
        \State ${A}_p\gets \argmax_{{A}\in\mathcal{P}}
               \left[\max\left(|\grad_p(\tilde{\gamma}, A)|, 
                               |\grad_p(-\tilde{\gamma}, A)|\right)\right]$
        \State Update $\vec{\gamma}_p\gets(\gamma_p, \vec{\gamma}_{p-1})$;
                       $\vec{\beta}_p\gets(\beta_p,\vec{\beta}_{p-1})$
        \State Add ${U}_p(\vec{\beta}_p,\vec{\gamma}_p) \gets 
                         e^{-i\beta_p{A}_p} e^{-i\gamma_p{H}} 
                         U_{p-1}(\vec{\beta}_{p-1},\vec{\gamma}_{p-1})$
        \EndIf
        \State Optimize params $\vec{\beta}_p^\star, \vec{\gamma}_p^\star 
                               \gets \argmax_{\vec{\beta}_p,\vec{\gamma}_p}
                               [E_p(\vec{\beta}_p,\vec{\gamma}_p)]$
        \State Set bound $\bound_p \gets E_p(\vec{\beta}_p^\star, 
                                             \vec{\gamma}_p^\star)$
        \If{$p = P$ \textbf{or} 
            $|\bound_{p-1}-\bound_{p}|<\varepsilon$}
            converged $\gets$ True
        \EndIf
        \EndWhile
    \State Sample bit strings from $\ket{\Psi_p^\star}$ and compute $\bound_p$
    \State \textbf{Return} bit strings, $\bound_p$, 
                           circuit $U_p$,
                           params $\vec{\beta}_p^\star$, $\vec{\gamma}_p^\star$
    \end{algorithmic}
\end{algorithm}

\section{Benchmarking}

In this section, we benchmark Dynamic- and standard ADAPT-QAOA in numerical 
simulations.
Our investigation will  demonstrate that Dynamic-ADAPT-QAOA can remove redundant 
components from the ansatz circuits of standard ADAPT-QAOA.
We show that this leads to a reduced CNOT count and an increased noise 
resilience.

\subsection{Benchmarking methodology}
\label{sec:benchmarking_methodology}

\begin{figure}
    \begin{center}
\includegraphics[scale=0.25]{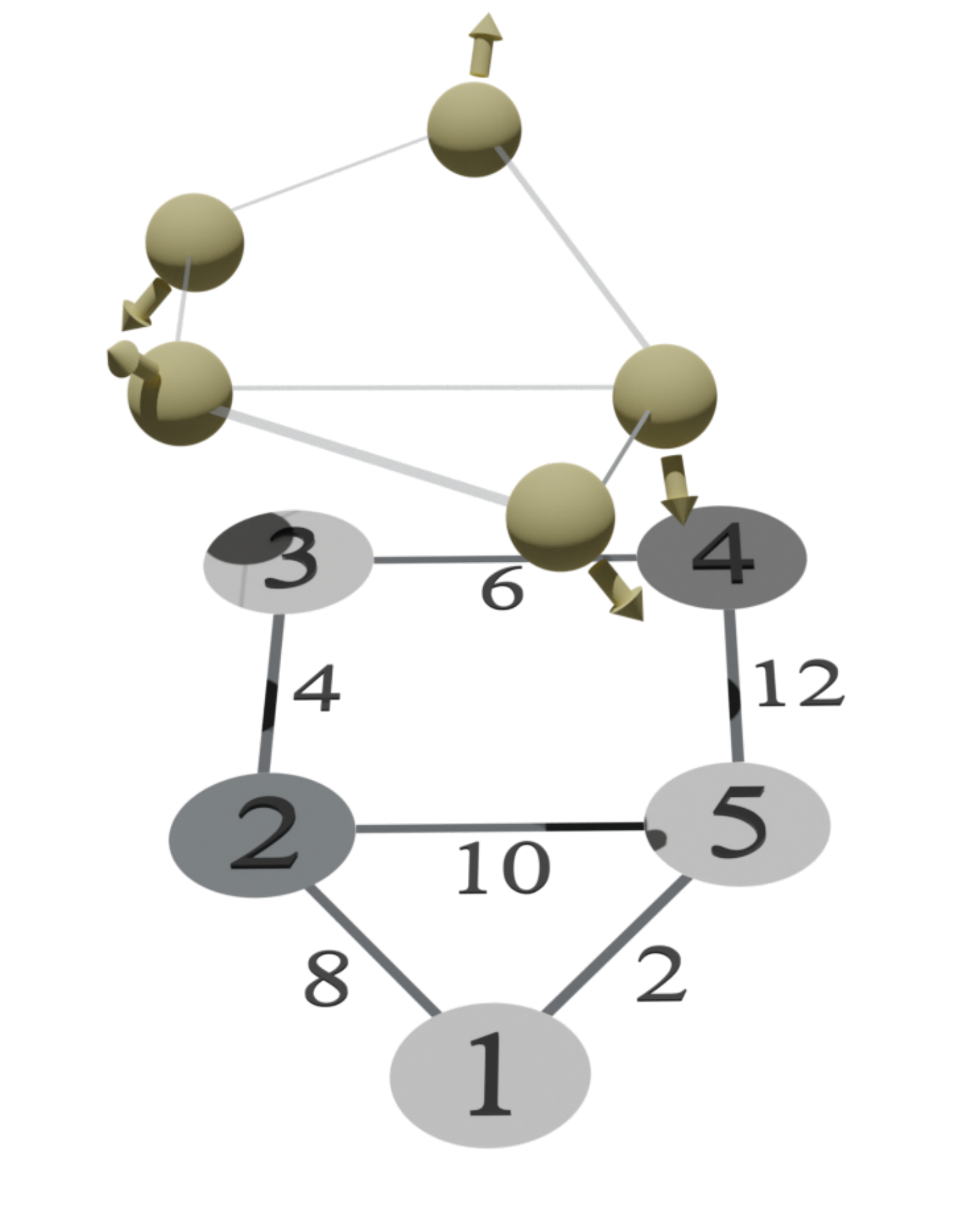}
    \end{center}
    \caption{Diagramatic representation of a 5-vertex weighted graph. The vertices are 
labelled 
    1-5. The weights are shown next to the corresponding edges. The 
partition
    resulting in a Max-Cut, $(135)(24)$, is depicted using different shades of gray. 
The Max-Cut value is $40$. Directly above the graph we illustrate how the 
problem maps onto a qubit system. The qubits' spins point 
in different vertical half-planes, corresponding to 
which set of the Max-Cut partition they are in.
    }\label{fig:maxcut}
\end{figure}

\textit{Max-Cut:---}In what follows, we benchmark ADAPT-QAOAs on random instances 
of  weighted Max-Cut problems.
Consider allocating weights to the edges of an $N$-vertex graph. In this work, we consider complete, i.e., fully connected, graphs. The edge weights between vertices $i\in N$ and $j \in N$ form a
real symmetric matrix $W_{ij}$ with zeros on its  diagonal. A binary vector $\vec{b}\in\{0,1\}^N$ defines a \textit{cut}, a  splitting of all vertices into two disjoint sets. A cut value is defined as the sum of edge weights between the two partitions:
\begin{align}
 \label{eq:CutValue}
 V(\vec{b}) = \sum_{i,j=1}^{N} W_{ij} b_i (1-b_j).
\end{align}
The weighted Max-Cut problem is to find the binary vector $\vec{b}^{\star}$ that maximizes the cut value: $\vec{b}^{\star} = \argmax_{\vec{b}}  V(\vec{b})$.
$\vec{b}^{\star}$ corresponds to the optimal partition, which yields the maximal cut value $\Vmax = V(\vec{b})^{\star}$.
By mapping binary variables $b_i=(1+z_i)/2$ to the eigenvalues $z_i \in \{-1,1\}$ 
of $Z_i$, the weighted Max-Cut problem becomes equivalent to finding the 
ground state of the Ising model, Eq.~\eqref{eq:cost_ham}.
We create random Max-Cut instances by uniformly sampling edge weights 
$W_{ij}\in[0,1]$.
This is known to generate NP-hard problems \cite{Commander2009,np_hard_maxcut}.
For a visualization of Max-Cut, see Fig.~\ref{fig:maxcut}.

\textit{Approximation ratio:---}Our benchmarks compare the average performance of three algorithms:\ 
Dynamic- and standard ADAPT-QAOA, as well as the classical, 
polynomial-time approximation algorithm by Goemans and Williamson 
(GW).
Rather than solving Max-Cut exactly, all three algorithms sample a collection 
of bit-strings \cite{williamson_shmoys_2011}.
This leads to a distribution of approximate cut values, Eq.~\eqref{eq:CutValue}, 
with average cut-values $\VD$, $\VS$, and $\VGW$, respectively.
Algorithms providing a higher average cut value tend to provide better-quality 
solutions.
Further, normalizing the average cut value by the maximal achievable value 
$\Vmax$ allows for averaging various instances of Max-Cut.
This defines our key performance metric---the average approximation ratio:
\begin{equation}
    \AD  \equiv \frac{\VD}{\Vmax}\text{, }
    \AS  \equiv \frac{\VS}{\Vmax}\text{, and }
    \AGW \equiv \frac{\VGW}{\Vmax}.
\end{equation}
The GW algorithm is the classical, polynomial-time algorithm that achieves the best worst-case approximation ratio:  $\AGW >
87.8 \ldots \%$ \cite{goemans_williamson}.
Below, we will compare $\AGW$ to numerically computed values of $\AD$ 
and $\AS$. In our simulations, we average the results over 100 random instances of the Max-Cut problem. In real applications of QAOA, one would return the cut corresponding to the sampled bit string with minimum cost, not the average. However, in the small problem sizes studied here, the final wavefunction has substantial overlap with all bit strings. Thus, for a relatively small number of shots the true solution will always be obtained. Therefore, we compare the average approximation ratios. Further, we emphasize that our comparison between QAOAs and the GW algorithm focuses on the final-results, i.e., average approximation ratios, not their computational time complexity.

\textit{Simulations:---}To assess the approximation ratios of Dynamic- and 
standard ADAPT-QAOA in the presence of noise, we use full density-matrix 
simulations, as previously described in Ref.~\cite{dalton2022variational}.
First, the unitaries in Eq.~\eqref{eq:CircuitUpdate} are compiled to 
standard circuit representations \cite{mixer_unitary_implementations}.
To simulate the effect of noise, we work with density matrices. In the evolution of the quantum states, we apply a depolarizing channel after each CNOT gate:
\begin{align}
    \mathcal{D}(i,\pgate)[\rho]\coloneq(1-\pgate)\rho +  
\frac{\pgate}{3}\sum_{\sigma_i}\sigma_i\rho\sigma_i.
\end{align}
Here, $\rho$ is the density matrix prior to the CNOT gate, $i$ denotes 
the target qubit of the CNOT gate, $\pgate \in [0,1]$ denotes the 
gate-error probability, and the $\sigma_i$-summation is over the three 
Pauli matrices acting on qubit $i$.
Owing to the diverse nature of current quantum hardware, a noise model cannot be both platform agnostic and realistically detailed.
Nevertheless, our noise model captures the depolarizing effect of two-qubit gates, which is the dominant noise source across several platforms \cite{nisq_noise_characterization, two_qubit_dominates}.
We deem our model a reasonably hardware-agnostic compromise, which should be 
sufficient to assess fundamental quantitative features.

Since full density-matrix simulations require extensive computing time, we 
apply an approximation similar to that outlined in Ref.~\cite{dalton2022variational}.
In more detail, we simulate ADAPT-QAOAs by growing their ansatz 
circuits in the absence of noise. We store the optimal ansatz circuits $U_p$ at 
each iteration step $p$.
Subsequently, we investigate the effect of noise by simulating the pre-optimized 
circuit $U_p$ at various noise levels $\pgate$ on our density matrix simulator.
As demonstrated in App.~\ref{appendix:noisy_growth}, the noiseless-growth approximation has little effect on our 
results. 

\textit{Parameters:---}Before presenting our findings, we specify 
the hyperparameters used in our simulations. By setting $\varepsilon =  
0$, we ensure that the convergence criterion corresponds to having reached a 
certain circuit depth. The depth is determined by the number of iterations, 
which we set to $P = 12$. For Dynamic-ADAPT-QAOA, the cost-unitary 
offset (see Algorithm \ref{alg:dynamic_adapt_qaoa}) was set to 
$\tilde{\gamma}=0.1$, following the settings used in \cite{adapt_qaoa}.
In Algorithm \ref{alg:dynamic_adapt_qaoa},  $\delta_1 > 0$ would mitigate 
some experimental errors in the identification of a local minimum where, in 
ideal scenarios, $C_p=0$. Similarly, $\delta_2 > 0 $ would mitigate some 
experimental errors in establishing whether $B_p\cdot D_p$ is positive. In our 
simulations, we set $\delta_1 = 0$. To emulate practical implementations, we 
choose $\delta_2\in(0, \, 10^{-4})$ after performing a hyperparameter 
search for each separate graph.

\subsection{Vanishing cost parameters}
\label{sec:vanish_cost_param}

\begin{figure}
    \includegraphics[width=\columnwidth]{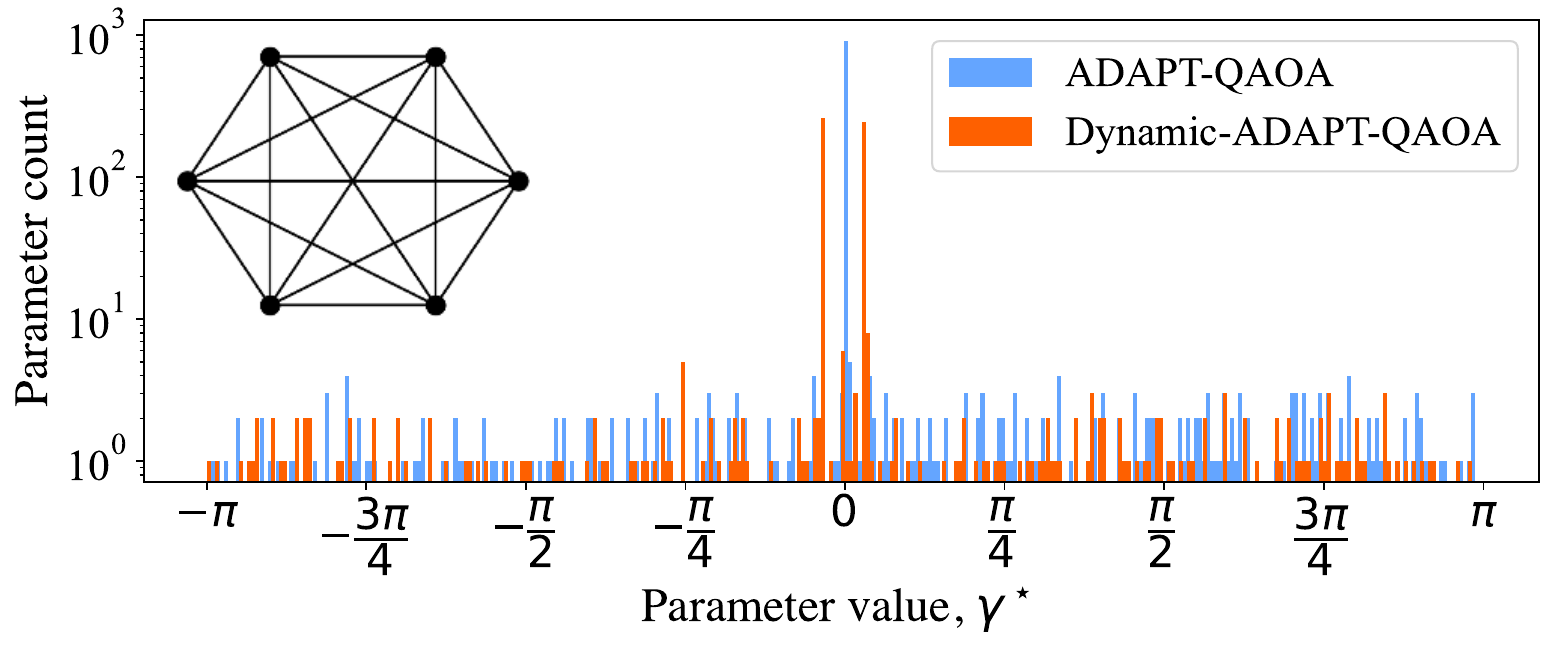}
    \caption{A histogram of optimized circuit parameters 
$\gamma_p^\star$, taken from the cost unitaries from all layers of the ansatz 
circuits grown with Dynamic- and standard ADAPT-QAOA. The data were acquired 
in noiseless simulations of 100 instances of Max-Cut on 
6-vertex graphs. The algorithms were run until a maximum circuit depth 
of $P=12$.}
\label{fig:histogram}
\end{figure}

As mentioned in Sec.~\ref{sec:dADAPT}, our motivation to develop 
Dynamic-ADAPT-QAOA stems from the observation that standard ADAPT-QAOA appends 
cost unitaries to the quantum circuit in cases where they do not lead to any 
significant improvement in convergence. In Figure \ref{fig:histogram}, we show 
data which support this conclusion. The histogram of optimal cost 
parameters $\gamma^\star$ of standard ADAPT-QAOA exhibits a well-defined peak 
at $\gamma^\star=0$. A majority ($\approx 70\%$) of the cost unitaries 
do not contribute to the algorithm's convergence. This peak is absent in 
the corresponding histogram for Dynamic-ADAPT-QAOA:  Our algorithm 
successfully removes redundant cost unitaries from the ansatz 
circuits.

\subsection{Benchmarking the CNOT-count reduction}
\label{sec:reducing_cnot_count}

\begin{figure}
    \includegraphics[width=\columnwidth]{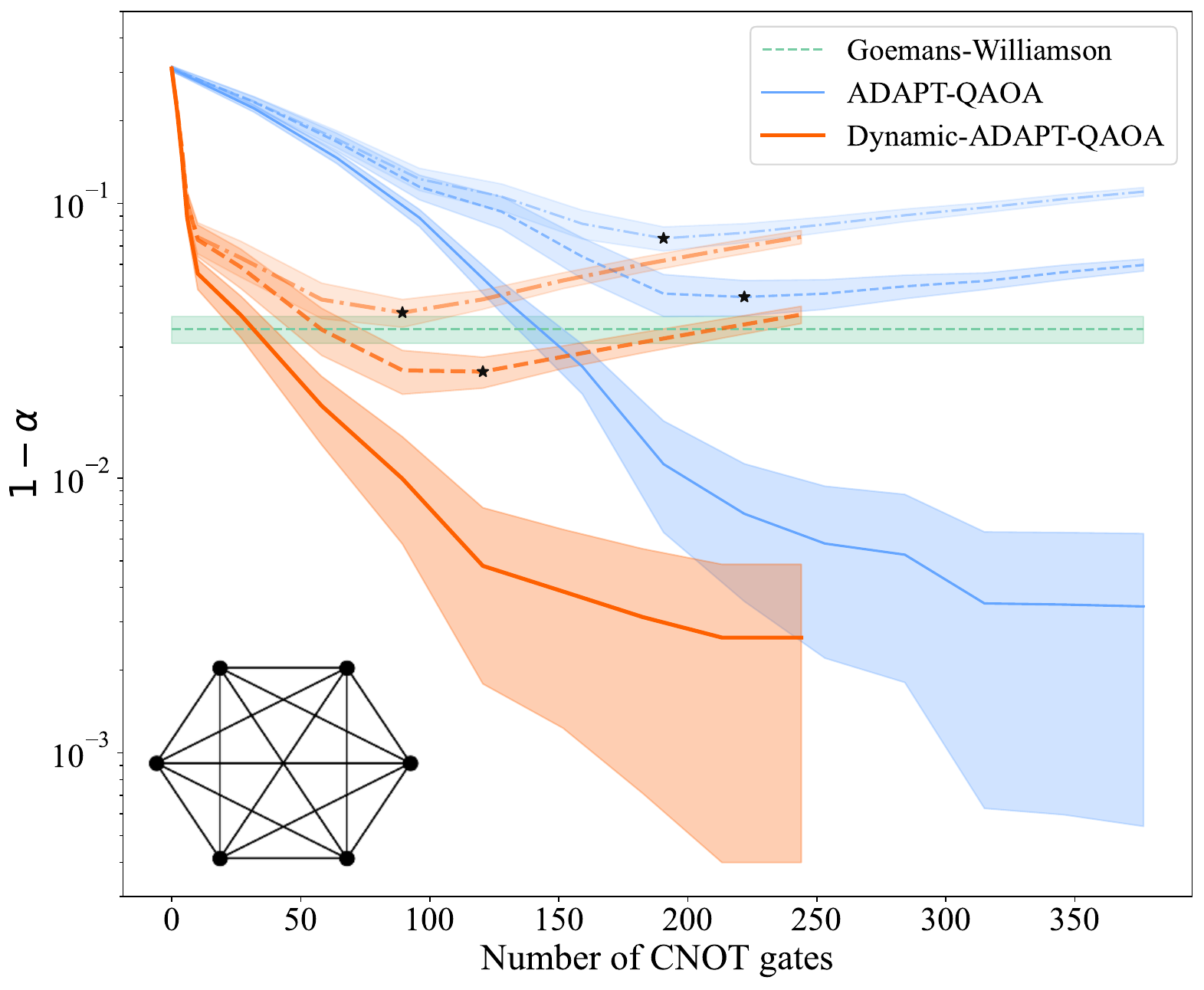}
    \caption{Convergence curves for Dynamic- and standard ADAPT-QAOA, applied to 6-vertex 
    complete graphs, with and 
without noise. $1-\alpha$ is plotted as a function of the number of CNOT 
gates present in the ansatz circuits $U_{P}$. The dashed horizontal 
curve corresponds to the classical GW algorithm. The shaded regions correspond 
to the $95\%$ confidence intervals. The convergence curves for three gate-error 
probabilities are shown: $\pgate=0.0\%, 0.122\%,$ and $0.263\%$. These are 
depicted using solid, dashed, and dash-dotted line styles, respectively. 
Stars indicate the maximally attainable approximation ratio 
$\alpha^\star$.
    }\label{fig:noisy_conv_curve}
\end{figure}

Now, we show that Dynamic-ADAPT-QAOA significantly reduces the 
number of CNOT gates needed to reach a certain algorithmic precision. In 
Section \ref{sec:algos}, we described how Dynamic-ADAPT-QAOA prunes unnecessary 
circuit elements.
To investigate the effect on the CNOT count, we consider how 
the approximation ratio $\alpha$, averaged over 100 instances of 
Max-Cut, improves as the algorithm 
grows the quantum circuit.
Our results are shown in FIG.~\ref{fig:noisy_conv_curve}. We plot data from both noiseless and noisy simulations of Dynamic- and standard ADAPT-QAOA.
In both scenarios,  Dynamic-ADAPT-QAOA uses significantly fewer 
CNOT gates to reach a fixed average approximation ratio.  For a fixed gate-error 
probability this CNOT reduction allows Dynamic-ADAPT-QAOA to calculate more 
accurate approximation ratios than standard ADAPT-QAOA. In noiseless 
simulations, we see that Dynamic-ADAPT-QAOA needs approximately $80\%$ fewer 
CNOT gates than ADAPT-QAOA to calculate average approximation ratios that outperform 
those achievable with the classical GW algorithm for 6-vertex complete graphs. 
Moreover, at a gate-error probability of $\pgate=0.122\%$, the 
Dynamic-ADAPT-QAOA can achieve better average approximation ratios than the GW algorithm, whilst the standard ADAPT-QAOA cannot. In the next section, we widen our 
analysis of how noise affects the quantum algorithms' achieved approximation ratios.
\begin{figure}
    \includegraphics[width=\columnwidth]{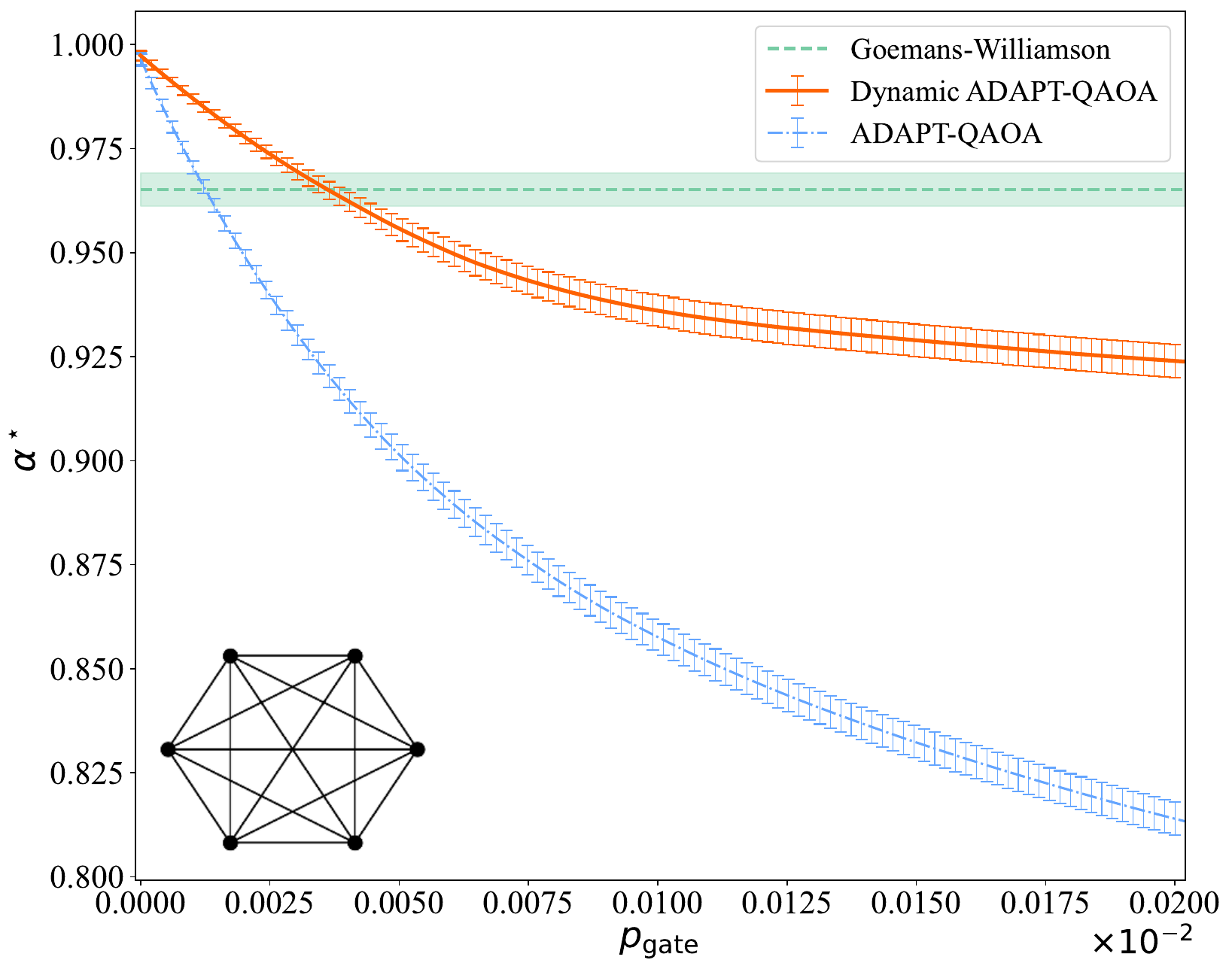}
    \caption{
Best attainable approximation ratio $\alpha^\star$ as a 
function of the gate-error probability $\pgate$. The data were acquired 
in noisy simulations of 6-vertex graphs. The error bars show the 
standard error in the mean approximation ratio. The dashed curve 
corresponds to the classical GW algorithm. The shaded regions correspond 
to the $95\%$ confidence intervals.}
    \label{fig:ratio_vs_error}
\end{figure}

\subsection{Benchmarking the noise resilience}
\label{sec:noise_resilience}

In this section, we analyze how noise affects the quality of approximation ratios of Dynamic- 
and standard ADAPT-QAOA.

The convergence curves presented in Fig.~\ref{fig:noisy_conv_curve} show that 
increasing the gate-error probability $\pgate$ worsens the best 
attainable average approximation ratio $\alpha^{\star}$.
More specifically, as ADAPT-QAOA grows the circuit (leading to an increase of 
CNOT gates on the abscissa) the approximation ratio improves initially.
However, as the circuit acquires more CNOT-gates, the effect of noise starts to 
dominate, leading to a subsequent deterioration of the approximation ratio.
This causes the characteristic ``smirk'' shape of the convergence curves in 
Fig.~\ref{fig:noisy_conv_curve}.
The dip of each convergence curve marks the best attainable average approximation 
ratio $\alpha^\star$ at a certain gate-error probability $\pgate$.

\begin{figure}
    \includegraphics[width=\columnwidth]{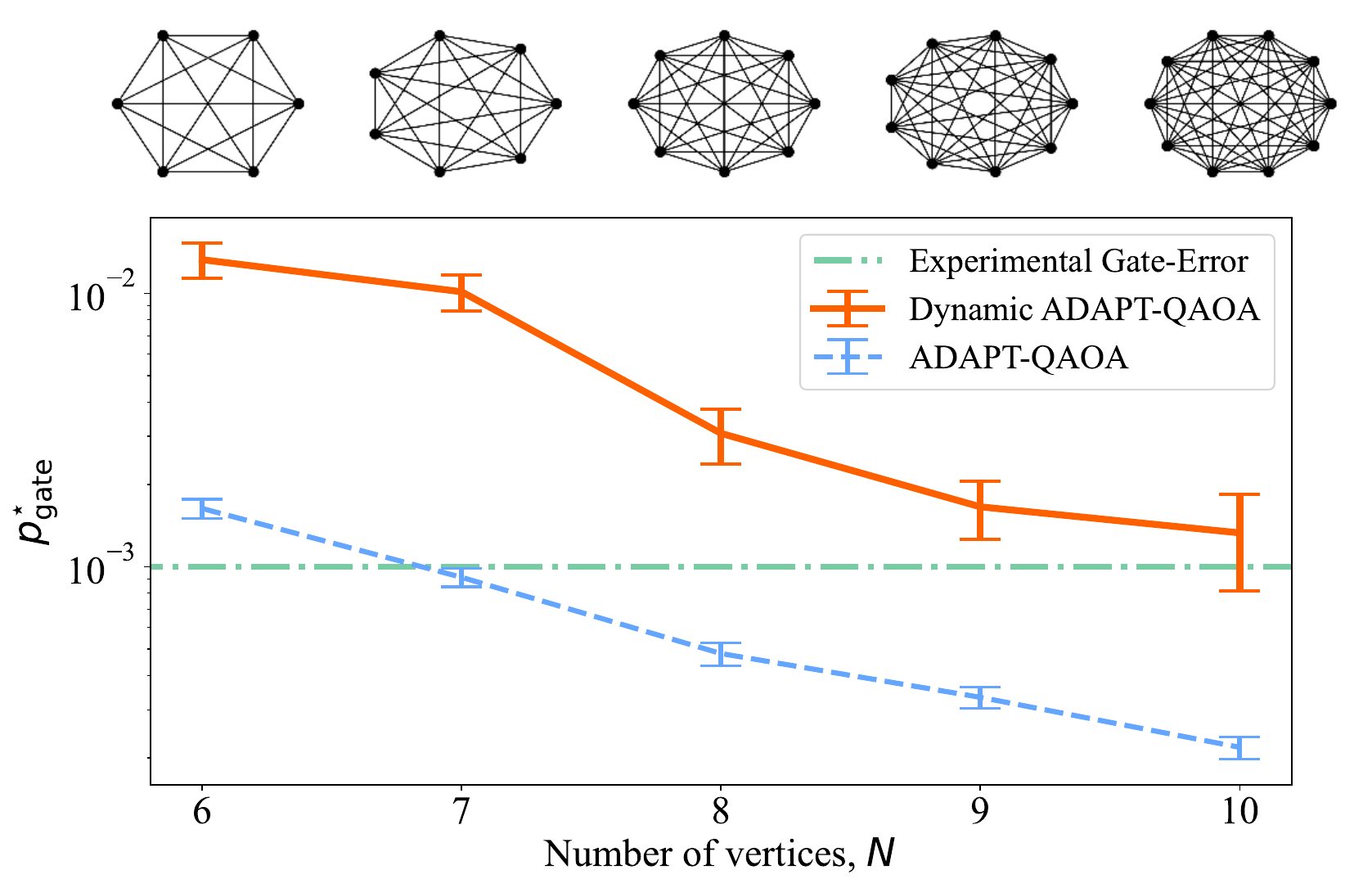}
\caption{$\pgate^\star$  with respect to different graph sizes. At 
gate-error probabilities below $\pgate^\star$  the quantum algorithms 
outperform the solution quality of the classical GW algorithm. The 
horizontal line shows the experimentally-achieved two-qubit gate-error 
probability in state-of-the-art superconducting hardware \cite{ding2023highfidelity}.
The error bars show the standard error.}
    \label{fig:main_result}
\end{figure}

Figure \ref{fig:noisy_conv_curve}  indicates that  Dynamic-ADAPT-QAOA 
outperforms the solution quality of standard ADAPT-QAOA in the 
presence of noise.
To quantify this benefit of our algorithm, we investigate 
$\alpha^\star$ as a function of $\pgate$ in FIG.~\ref{fig:ratio_vs_error}. For 
all values of $\pgate$,  Dynamic-ADAPT-QAOA calculates better 
approximation ratios than standard ADAPT-QAOA. Evidently, our algorithm
exhibits better noise resilience.\\
\indent As can be seen from the left-most portion of FIG.~\ref{fig:ratio_vs_error}, 
given sufficiently weak noise, both Dynamic- and standard ADAPT-QAOA can provide better average approximation ratios than the GW algorithm.
We now investigate the range of gate-error probabilities for which Dynamic- and 
standard ADAPT-QAOAs achieve such an improvement.
To this end, we define the gate-error probability $\pgate^\star$, below which the quantum algorithms achieve a better average approximation ratio than the GW algorithm.
In FIG.~\ref{fig:main_result}, we plot $\pgate^\star$ with respect to the number 
of graph vertices. Compared to standard ADAPT-QAOA, Dynamic-ADAPT-QAOA
can achieve a better max-cut approximation ratio than the classical GW 
algorithm at roughly an order of magnitude larger values of $\pgate^\star$. In 
particular, the critical probability at which Dynamic-ADAPT-QAOA 
achieves higher approximation ratios than the GW algorithm is 
$\pgate^\star=1.3\pm0.2\%$ for 6-vertex graphs and  $\pgate^\star=0.13\pm0.05\%$ 
for 10-vertex graphs.
Both  these values are well above achieved gate-error probabilities \cite{ding2023highfidelity}, implying that one may execute Dynamic-ADAPT-QAOA on existing hardware.
On the other hand, for standard ADAPT-QAOA, the critical probability is currently 
achievable only for graphs with less than 7 vertices.

\section{Discussion}
\label{sec:discussion}

We have introduced Dynamic-ADAPT-QAOA, a quantum algorithm for 
combinatorial optimization. Similar to the original ADAPT-QAOA 
algorithm, our algorithm variationally approximates the ground state of an 
Ising Hamiltonian. Thus, it can provide approximate solutions to NP 
problems. By dynamically assessing the importance of unitaries before they are 
added in the variationally grown algorithms, Dynamic-ADAPT-QAOA can operate 
with remarkably few CNOT gates.  Above, we benchmarked the average (as opposed to the worst-case) performance of our algorithm.
For example, in the idealized case of no noise, Dynamic-ADAPT-QAOA requires on average about 35 (350) CNOT gates to outperform the GW algorithm on 6-vertex (10-vertex) graphs. 
Moreover, we have shown that for graphs 
with $6-10$ vertices, Dynamic-ADAPT-QAOA can provide better average solutions than the GW algorithm, even in the 
presence of noise levels comparable with current state-of-the-art hardware 
\cite{ding2023highfidelity}. This should make Dynamic-ADAPT-QAOA an attractive candidate to showcase 
proof-of-principle  computations on NISQ hardware. 
Finally, we conclude this work with a few comments.

\textit{Other QAOAs:---}There are plenty of promising QAOA algorithms in the 
literature \cite{qaoa,symmetric_qaoa,grover_qaoa,qaoa_1,qaoa_2,qaoa_3,qaoa_4,qaoa_implementation_trapped_ion,fermionic_qaoa, Farhi22}.
However, this work focuses on ADAPT-QAOAs \cite{adapt_qaoa}---mainly due to 
their relatively shallow ansatz circuits. In the future, it would be of 
interest to expand the benchmarks of noise resilience to other types of QAOA. 

\textit{Other algorithms:---}This study 
focuses on investigating the utility of gate-based quantum computers for 
solving NP-problems. 
However, adiabatic quantum computers \cite{quantum_annealing,adiabatic_quantum_computation,manufactured_spins, 
quantum_adiabatic_evolution} and state-of-the-art annealing heuristics \cite{quantum_annealing_2,simulated_annealing_1,simulated_annealing_2,simulated_annealing_3,momentum_annealing}  can comfortably handle systems with up to $5$ thousand and $100$ thousand 
spins, respectively, most likely at a higher solution accuracy.
Moreover, other approximation algorithms \cite{Montanari19, Alaoui23} could also lead to high average solution accuracy.
This shows that QAOA still has a long way to go before reaching practical quantum advantage.\\
\indent\textit{Error mitigation:---}Applying error-mitigation techniques \cite{error_mitigation_1,error_mitigation_2,error_mitigation_3,error_mitigation_4} to boost 
expectation values would straightforwardly improve the approximation ratios of standard and  
Dynamic-ADAPT-QAOA, see App.~\ref{app:error_mitigation}. 
However, to the best of our knowledge, error-mitigation methods have never been used to improve the underlying bit-strings.
Consequently, error-mitigation methods would not improve the cut value provided by the experimentally accessible bit-strings.
An interesting direction of future research is to consider how error-mitigation techniques could be used to improve not only the cut value, but also the bit-strings provided by a QAOA.

\textbf{Acknowledgements:} We thank Kieran Dalton, Yordan 
Yordanov, Bobak Kiani, Nicholas Mayhall, Sophia Economou,  Edwin Barnes,  and members of the Hitachi QI team for useful discussions.

\pagebreak
\bibliographystyle{unsrt}
\bibliography{refs}

\onecolumngrid

\newpage
\clearpage

\appendix

\section{CNOT gate count in cost unitaries}
\label{app:CNOT_count}

In this appendix we show that the number of CNOT gates included in implementations
of the cost unitaries in QAOA scales quadratically with the number of vertices for 
complete graphs. This forms part of the motivation for our development 
of Dynamic-ADAPT-QAOA.\\
\indent The general form of a parameterized cost unitary is $e^{-i\gamma H}$,
where the Ising Hamiltonian is defined in Eq.~(\ref{eq:cost_ham}). For complete graphs,
$H$ is a summation of $\mathcal{O}(N^2)$ two-qubit Pauli-strings, where $N$
equals the number of vertices. As all of these Pauli-strings are of the same general form,
namely with two $z$-measurements, they all commute with one another. Hence, we can
re-write the cost unitary as a product of $\mathcal{O}(N^2)$ RZZ gates:
\begin{align}
    e^{-i\gamma H} = \prod_{i,j=1}^N \exp\left(-i\frac{\gamma W_{ij}}{4}Z_iZ_j\right).
\end{align}
Each of these RZZ gates can be implemented using two CNOT gates. Therefore, 
we conclude that each cost unitary included in the parameterized unitary of the
algorithm contributes $\mathcal{O}(N^2)$ CNOT gates to the quantum circuit decomposition.

\section{Theoretical aspects underlying Dynamic-ADAPT-QAOA}

\label{Appendix:EnergyMinimaAnalysis}

In this appendix we provide further details on the analysis of minima in the 
energy variation, Eq.~\eqref{eq:two_dim_func_opt_mixer}.

\subsection{Splitting of cost Hamiltonians}

To begin with, we note that any cost Hamiltonian $H$, Eq.~\eqref{eq:cost_ham}, 
can be decomposed into two parts, $H_{-}$ and $H_{+}$, which commute and 
anti-commute with a given Pauli-string mixer $A$, respectively.
To show this, denote the qubit indices as a vertex set 
$ \mathcal{V} := \left\{1,...,N\right\} $
with the corresponding edge set given as
$
 \mathcal{E} 
 = \left\{ (i,j) \in\mathcal{V}\times\mathcal{V} \,|\, i<j\right\}.
$
This allows for writing the cost Hamiltonian, Eq.~\eqref{eq:cost_ham}, as
\begin{align}
 H = \frac{1}{2}\sum_{(i,j)\in\mathcal{E}} W_{ij} Z_i Z_j.
\end{align}
To split the Hamiltonian, we split the edge set $\mathcal{E}$ into two 
disjoint subsets $\mathcal{E}_{-}$ and $\mathcal{E}_{+}$, such that 
\begin{align}
 \mathcal{E} = \mathcal{E}_{-} \cup \mathcal{E}_{+} 
 \quad \text{and} \quad
 \mathcal{E}_{-} \cap \mathcal{E}_{+} = \emptyset.
\end{align}
This defines two Hamiltonians
\begin{align}
 H_{-} = \frac{1}{2}\sum_{(i,j)\in\mathcal{E}_{-}} W_{ij} Z_i Z_j 
 \quad \text{and} \quad
 H_{+} = \frac{1}{2}\sum_{(i,j)\in\mathcal{E}_{+}} W_{ij} Z_i Z_j,
\end{align}
such that
\begin{align}
 H = H_{-} + H_{+}.
\end{align}
For later reference, we note that $H$, $H_{-}$, and $H_{+}$ commute.

For mixer Hamiltonians $A\in\pool$, which are Pauli-strings of length one, 
acting on qubit $n=1,...,N$ via $X_n$ or $Y_n$, we partition the edge set of 
$H$ as follows
\begin{align}
\mathcal{E}_{-}(n)
&=\left\{(i,j)\in\mathcal{E} \quad|\quad \text{if neither $i$ nor $j$ is $n$} 
\right\},\\
\mathcal{E}_{+}(n)
&=\left\{(i,j)\in\mathcal{E} \quad|\quad \text{if either $i$ or $j$ is $n$}
\right\}.
\end{align}
For mixer Hamiltonians $A\in\pool$, which are Pauli-strings of length two, 
acting on qubits $n,n'=1,...,N$, we partition the edge sets as follows:\
(i) If the Pauli string takes the form $Z_n \cdot Z_{n'}$, the partition of 
edges is trivial $\mathcal{E}_{-}=\mathcal{E}$ and 
$\mathcal{E}_{+}=\emptyset$.
(ii) If the Pauli string is of the form $X_{n} \cdot Z_{n'}$, we use the 
partition of single Pauli strings acting on $n$.
(iii) If the Pauli-string is of the form $X_n \cdot X_{n'}$, $X_n \cdot 
Y_{n'}$, or $Y_n \cdot Y_{n'}$, respectively, we partition the edge set of $H$ 
as follows
\begin{align}
\mathcal{E}_{-}(n, n')
&=\left\{(i,j)\in\mathcal{E} \quad|\quad \text{if $i$ and $j$ differ from 
$n$ and $n'$, or if $i=n$ and $j=n'$, or $j=n$ and $i=n'$} \right\},\\
\mathcal{E}_{+}(n, n')
&=\left\{(i,j)\in\mathcal{E} \quad|\quad \text{if either $i$ or $j$ is 
equal to $n$ or $n'$, but not both}
\right\}.
\end{align}
These partitions ensure that $H_{-}$ and $H_{+}$ commute or anti-commute with 
Pauli-strings $A$ of length one or two, respectively.

\subsection{Energy variation}

Next, we analyze the energy variation, Eq.~\eqref{eq:two_dim_func_opt_mixer}, 
with respect to a mixer $A$, as defined by Eq.~\eqref{eq:state_variation} and 
Eq.~\eqref{eq:two_dim_func}:
\begin{align}
    \label{eq:energy_fluctuation}
 \fluc_p(\beta_p,\gamma_p; A) 
 = \Bra{\Psi_{p-1}^{\star}} e^{i\gamma_p H} e^{i\beta_p A} H 
                            e^{-i\beta_p A}e^{-i\gamma_p H} 
    \Ket{\Psi_{p-1}^{\star}}.
\end{align}
To analyze this expression further we rewrite it as
\begin{align}
    \label{eq:energy_fluctuation_rewriteI}
 \fluc_p(\beta_p,\gamma_p; A) 
 = \Bra{\Psi_{p-1}^{\star}} H_{-} \Ket{\Psi_{p-1}^{\star}}
 + \cos(2\beta_p) \Bra{\Psi_{p-1}^{\star}} H_{+} \Ket{\Psi_{p-1}^{\star}}
 + \sin(2\beta_p) \Bra{\Psi_{p-1}^{\star}} iA H_{+} e^{-i2\gamma_p H_{+}} 
\Ket{\Psi_{p-1}^{\star}}
\end{align}
To derive this expression we used the following properties of Pauli-strings $A$
\begin{align}
 \text{id} = A^2 \quad\text{and}\quad
 e^{-i\beta_p A} = \text{id} \cos(\beta_p) + i A \sin(\beta_p).
\end{align}
Moreover, we use that $H = H_{-} + H_{+}$ and the established commutation 
relations implying
\begin{align}
    H_{-} A = A H_{-} \quad\and\quad H_{+} A = -A H_{+},
\end{align}
as well as
\begin{align}
    e^{-i\gamma_p H} = e^{-i\gamma_p H_{+}}e^{-i\gamma_p H_{-}} = 
    e^{-i\gamma_p H_{+}}e^{-i\gamma_p H_{-}}. 
\end{align}
Next, defining the following quantities
\begin{align}
  E_0^{-} &= \Bra{\Psi_{p-1}^{\star}} H_{-} \Ket{\Psi_{p-1}^{\star}}, \\
  E_0^{+} &= \Bra{\Psi_{p-1}^{\star}} H_{+} \Ket{\Psi_{p-1}^{\star}}, \\
  B(\gamma_p) 
  &= \Bra{\Psi_{p-1}^{\star}} iA H_{+} e^{-i2\gamma_p H_{+}} 
                                           \Ket{\Psi_{p-1}^{\star}},\\
  C(\gamma_p) &= \Bra{\Psi_{p-1}^{\star}} A H_{+}^2 e^{-i2\gamma_p H_{+}} 
                                           \Ket{\Psi_{p-1}^{\star}}
               = \frac{1}{2}\partial_{\gamma_p}B(\gamma_p),\\
  D(\gamma_p) &= \Bra{\Psi_{p-1}^{\star}} iA H_{+}^3 e^{-i2\gamma_p H_{+}} 
                                           \Ket{\Psi_{p-1}^{\star}}
               = -\frac{1}{2}\partial_{\gamma_p}C(\gamma_p),
\end{align}
one rewrites the energy fluctuation, Eq.~\eqref{eq:energy_fluctuation_rewriteI}, 
as
\begin{align}
    \label{eq:energy_fluctuation_rewriteII}
 \fluc_p(\beta_p,\gamma_p; A) 
 = E_0^{-} + E_0^{+} \cos(2\beta_p) + \sin(2\beta_p) B(\gamma_p)
\end{align}

\subsection{Gradients, stationary points and Hessian of the energy variation}

The gradient of the energy fluctuation are then given as 
\begin{align}
 \label{eq:energy_fluctuation_gradient}
  \partial_{\gamma_p}\fluc_p(\beta_p,\gamma_p; A) 
  &= 2 \sin(2\beta_p) C(\gamma_p)\\
  \partial_{\beta_p}\fluc_p(\beta_p,\gamma_p; A) 
  &= 2 \left[-\sin(2\beta_p) E_{0}^{-} + \cos(2\beta_p) B(\gamma_p)\right]
\end{align}
Evaluating the gradient at $\gamma_p=0,\beta_p=0$ gives
\begin{align}
 |\grad_p(A)| = |2B(0)| =  \left|\Bra{\Psi_{p-1}^{\star}} iA H_{+}  
                                           \Ket{\Psi_{p-1}^{\star}}\right|.
\end{align}
We further note that the stationary points $\bar{\beta}_p, \bar{\gamma}_p$ 
where the gradient vanishes, fulfill the following conditions
\begin{align}
  0 = \sin(2\bar{\beta}_p) C(\bar{\gamma}_p)\quad\text{and}\quad
  \sin(2\bar{\beta}_p) E_{0}^{-} = \cos(2\bar{\beta}_p) B(\bar{\gamma}_p).
\end{align}
This implies stationary points of two types:
\begin{align}
 &\text{Type 1: }\sin(2\bar{\beta}_{p,1})=0 \quad\text{and}\quad 
B(\bar{\gamma}_{p,1}) = 0\\
 \label{eq:type2}
 &\text{Type 2:    } C(\bar{\gamma}_{p,2}) = 0 \quad\text{and}\quad 
E_0^{-}\sin(2\bar{\beta}_{p,2})=B(\bar{\gamma}_{p,2})\cos(2\bar{\beta}_{p,2}).
\end{align}

Further computing the second derivatives
\begin{align}
 \label{eq:energy_fluctuation_second_derivatives}
  \partial_{\gamma_p}^2\fluc_p(\beta_p,\gamma_p; A) 
  &= -4 \sin(2\beta_p) D(\gamma_p),\\
  \partial_{\beta_p}\partial_{\gamma_p}\fluc_p(\beta_p,\gamma_p; A) 
  &= 4 \cos(2\beta_p) C(\gamma_p),\\
  \partial_{\gamma_p}\partial_{\beta_p}\fluc_p(\beta_p,\gamma_p; A) 
  &= 4 \cos(2\beta_p) C(\gamma_p),\\
  \partial_{\beta_p}^2\fluc_p(\beta_p,\gamma_p; A) 
  &= -4 \left[\cos(2\beta_p) E_{0}^{-} + \sin(2\beta_p) B(\gamma_p)\right],
\end{align}
gives the Hessian
\begin{align}
 \label{eq:energy_fluctuation_Hessian}
 \text{Hess}(\beta_p, \gamma_p) = 
 \begin{pmatrix}
  -4 \sin(2\beta_p) D(\gamma_p) &  4 \cos(2\beta_p) C(\gamma_p) \\
   4 \cos(2\beta_p) C(\gamma_p) & -4 \left[\cos(2\beta_p) E_{0}^{-} 
                                  +\sin(2\beta_p) B(\gamma_p)\right]\\
 \end{pmatrix}.
\end{align}
The determinant of the Hessian is given as
\begin{align}
 \text{Det}\left[\text{Hess}(\beta_p,\gamma_p)\right]
    =16\left[\sin(2\beta_p)\cos(2\beta_p)D(\gamma_p)E_{0}^{-}
              +\sin^2(2\beta_p) B(\gamma_p)D(\gamma_p)
              -\cos^2(2\beta_p) C(\gamma_p) \right].
\end{align}
Evaluating the determinant at the stationary points results in
\begin{align}
 \text{Det}\left[\text{Hess}(\bar{\beta}_{p,1},\bar{\gamma}_{p,1})\right]
   =-16\cos^2(2\bar{\beta}_{p,1}) C^2(\bar{\gamma}_{p,1}) \quad\text{and}\quad
 \text{Det}\left[\text{Hess}(\bar{\beta}_{p,2},\bar{\gamma}_{p,2})\right]
   =16 B(\bar{\gamma}_{p,2})D(\bar{\gamma}_{p,2}).
\end{align}
The determinant of type 1 stationary points is always negative, indicating a 
positive and a negative eigenvalue of the Hessian matrix.
This implies that type 1 stationary points are saddle point of the energy 
variation. 
On the other hand, the determinant of type 2 stationary points is positive if
\begin{align}
 \label{eq:PositiveDeterminant}
 B(\bar{\gamma}_{p,2})D(\bar{\gamma}_{p,2})>0.
\end{align}
This implies that both eigenvalues of the Hessian matrix have the same sign, and 
thus indicates the existence of a minimum or a maximum of the energy variation.

Next, we analyze the trace of the Hessian, which is identical to the sum of its 
eigenvalues
\begin{align}
 \text{Tr}\left[\text{Hess}(\beta_p,\gamma_p)\right]
    &=-4\left[\sin(2\beta_p) D(\gamma_p)
              + \cos(2\beta_p) E_{0}^{-} + \sin(2\beta_p) B(\gamma_p) \right].
\end{align}
Evaluating the trace at type 2 stationary points, Eq.~\eqref{eq:type2}, and 
using its properties results in
\begin{align}
 \text{Tr}\left[\text{Hess}(\bar{\beta}_{p,2},\bar{\gamma}_{p,2})\right]
    &=-4\left\{\sin(2\bar{\beta}_{p,2}) 
              \left[D(\bar{\gamma}_{p,2}) + B(\bar{\gamma}_{p,2})\right]
              + \cos(2\bar{\beta}_{p,2}) E_{0}^{-} \right\},\\
    &=-4\sin(2\bar{\beta}_{p,2}) \left[
                                    D(\bar{\gamma})_{p,2}
                                    + B(\bar{\gamma})_{p,2}
                                    +\frac{(E_{0}^{-})^2}{B(\bar{\gamma}_{p,2})}
                                 \right].
\end{align}
To analyze the sign of the trace further, note that there are two distinct type-2 stationary points
\begin{align}
 \bar{\beta}_{p,2,-}\in[-\pi/2,0] 
 \quad\text{and}\quad 
 \bar{\beta}_{p,2,+}\in[0, \pi/2].
\end{align}
Depending on the sign of $D(\bar{\gamma})_{p,2} + 
B(\bar{\gamma})_{p,2}+(E_{0}^{-})^{2}/B(\bar{\gamma}_{p,2})$ one of these 
solutions will lead to a positive and the other one to a negative trace.
Thus, as long as Eq.~\eqref{eq:PositiveDeterminant} is fulfilled, the energy 
variation, will always have a unique minimum.

\subsection{Assessment criteria for Dynamic-ADAPT-QAOA}

Summarizing the aforementioned analysis, Eq.~\eqref{eq:type2} and 
Eq.~\eqref{eq:PositiveDeterminant} imply that the energy variation, 
Eq.~\eqref{eq:two_dim_func_opt_mixer}, has a unique minimum at 
$\bar{\gamma}_{p,2}=0$ if 
\begin{align}
 C(0)=0 \quad\text{and}\quad B(0)D(0)>0.
\end{align}
Substituting $A=A_p$ into these equations, allows for identifying $B(0)\equiv 
B_p$, $C(0)\equiv C_p$, and $D(0)\equiv D_p$ (as defined in 
Eq.~\eqref{eq:param_space_quantities}) and results in 
Condition~\eqref{eq:Minima}.
Dynamic-ADAPT-QAOA tests this condition to confirm whether the energy 
fluctuation has a unique minimum at $\bar{\gamma}_{p,2}=0$, indicating an 
optimal value at $\gamma_p^{\star}\approx0$.

\section{Comparison of Dynamic-ADAPT-QAOA versions}
\label{app:algo_subtleties}
\begin{figure}
    \centering
    \includegraphics[width=0.7\textwidth]{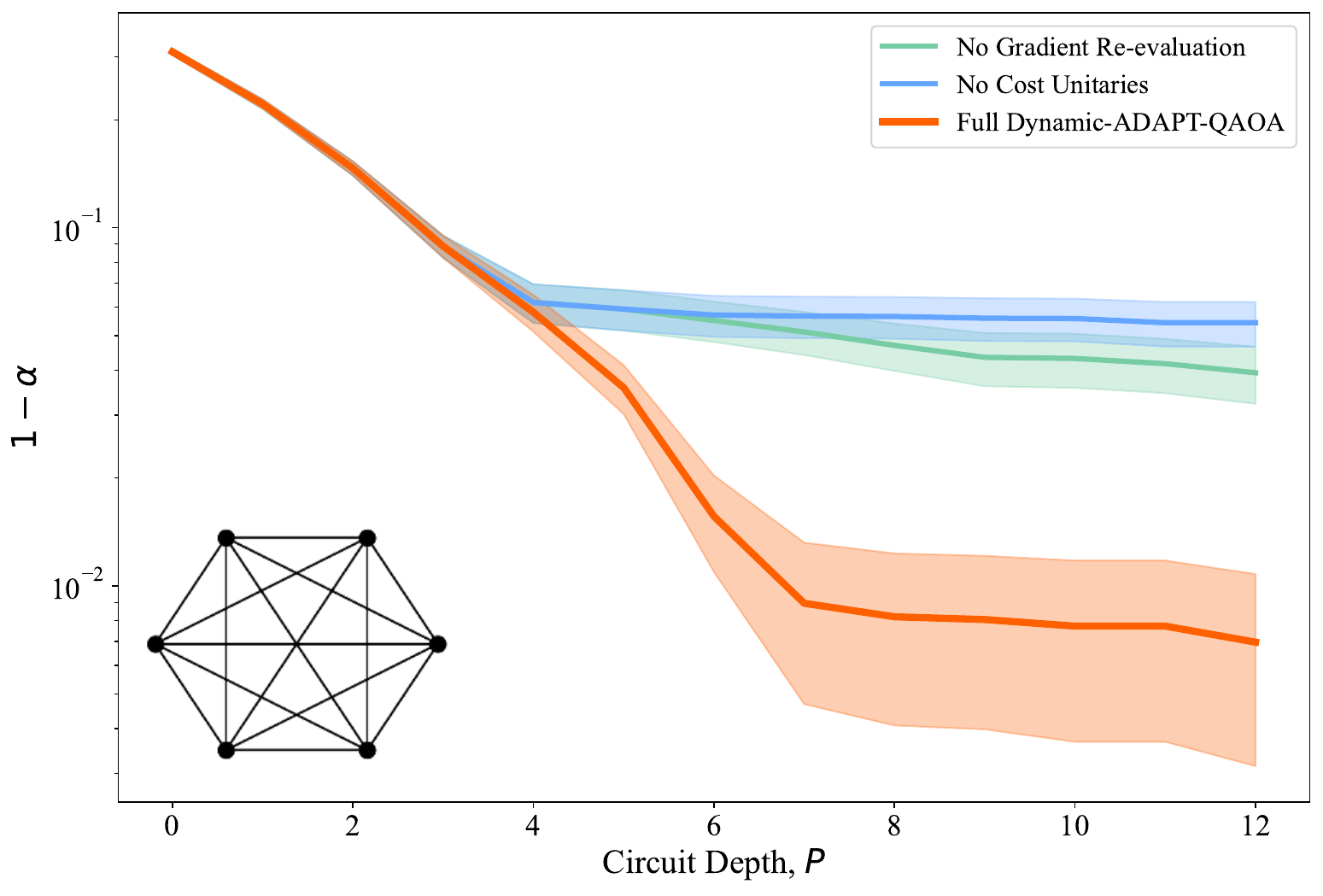}
    \caption{Convergence curves for three versions of Dynamic-ADAPT-QAOA, applied
    to 6-vertex complete graphs, without noise. $1-\alpha$ is plotted 
    as a function of the depth $P$ of the parameterized unitary $U_P$. The shaded regions 
    depict $95\%$ confidence intervals for the mean approximation ratio, averaged over 100 randomized graphs.
    }
    \label{fig:compare_dADAPT_versions}
\end{figure}

In this appendix we provide numerical evidence supporting the two details presented at the end
of Sec.~\ref{sec:dADAPT} related to the execution of Dynamic-ADAPT-QAOA. In particular, we
analyze the convergence of three versions of Dynamic-ADAPT-QAOA:
\begin{itemize}
    \item The full version which we consider in all other parts of this paper;
    \item A version in which we remove the cost unitaries from all layers of the quantum circuit;
    \item A version in which we do not re-evaluate the energy gradients $\grad_p(\gamma_p;A)$,
    and keep using the same mixer $A_p$ in the case where Condition (\ref{eq:Minima}) is not
    met.
\end{itemize}
The data is presented in FIG.~\ref{fig:compare_dADAPT_versions}. We find that the full version of Dynamic-ADAPT-QAOA
produces, on average, a much better approximation ratio when compared to the other two tested versions. We conclude
that the two aforementioned subtleties to our algorithm are justified, and 
contribute to its improved performance.

\section{Noiseless and noisy circuit growth}
\label{appendix:noisy_growth}

\begin{figure}
    \centering
    \includegraphics[width=0.7\textwidth]{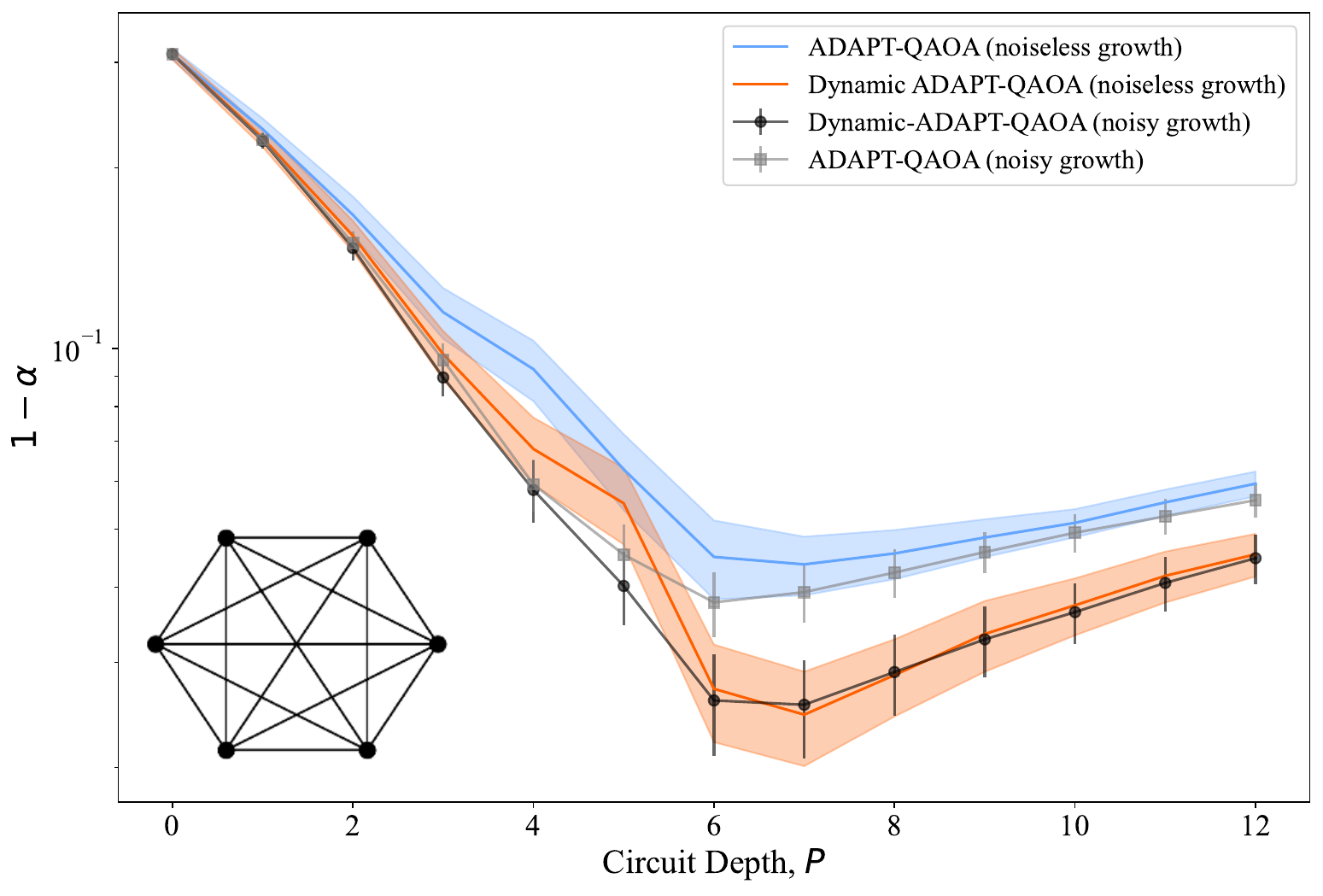}
    \caption{Convergence curves for Dynamic- and standard ADAPT-QAOA, applied
    to 6-vertex complete graphs, with noise. $1-\alpha$ is plotted 
    as a function of the depth $P$ of the parameterized unitary $U_P$. The data for both noiseless and noisy
    growth of the quantum circuit are presented. The error bars and shaded regions 
    depict the standard error in the mean approximation ratio, averaged over 100 randomized graphs.}
    \label{fig:noisy_growth}
\end{figure}

In this appendix we present numerical evidence supporting our approach to simulating the performance of Dynamic- and standard ADAPT-QAOA. In section \ref{sec:benchmarking_methodology} we mention two methods for analyzing the evolution of the approximation ratio achieved by the quantum algorithms. In one, the effects of noise are implemented during the circuit growth. This is representative of how the algorithm would be run in an experiment using real hardware. However, simulating this approach using density matrices requires large computing times, making it impractical. Therefore, we make use of the alternative method, used previously in the context of variational quantum eigensolvers \cite{dalton2022variational}. This alternative approach relies on the theoretical result that noise in variational quantum algorithms primarily flattens the parameter landscape, without altering its structure \cite{noise_flattening}.\\
\indent In particular, we grow the quantum circuits in the absence of noise, recording the optimal components and parameters found during each iteration. To investigate the effect of noise afterward, we simulate the pre-optimized circuits while including the respective noise channel where necessary. This greatly reduces the computational time required for running the quantum algorithms, making it the preferable method for numerical investigations.\\
\indent It is important to check whether growing the quantum circuits in the noiseless regime significantly alters the results of our analysis. To address this, we compare the two methods described above in the context of 6-vertex graphs. The resulting data is shown in FIG.~\ref{fig:noisy_growth} for a gate-error probability of $p_\text{gate}=0.122\%$. We find that both methods produce, on average, the same behaviour for either algorithm. This justifies our use of the less computationally time-consuming approach for circuit growth simulation.\\

\section{Error mitigation in Dynamic-ADAPT-QAOA}
\label{app:error_mitigation}

\begin{figure}
    \centering
    \includegraphics[width=0.7\textwidth]{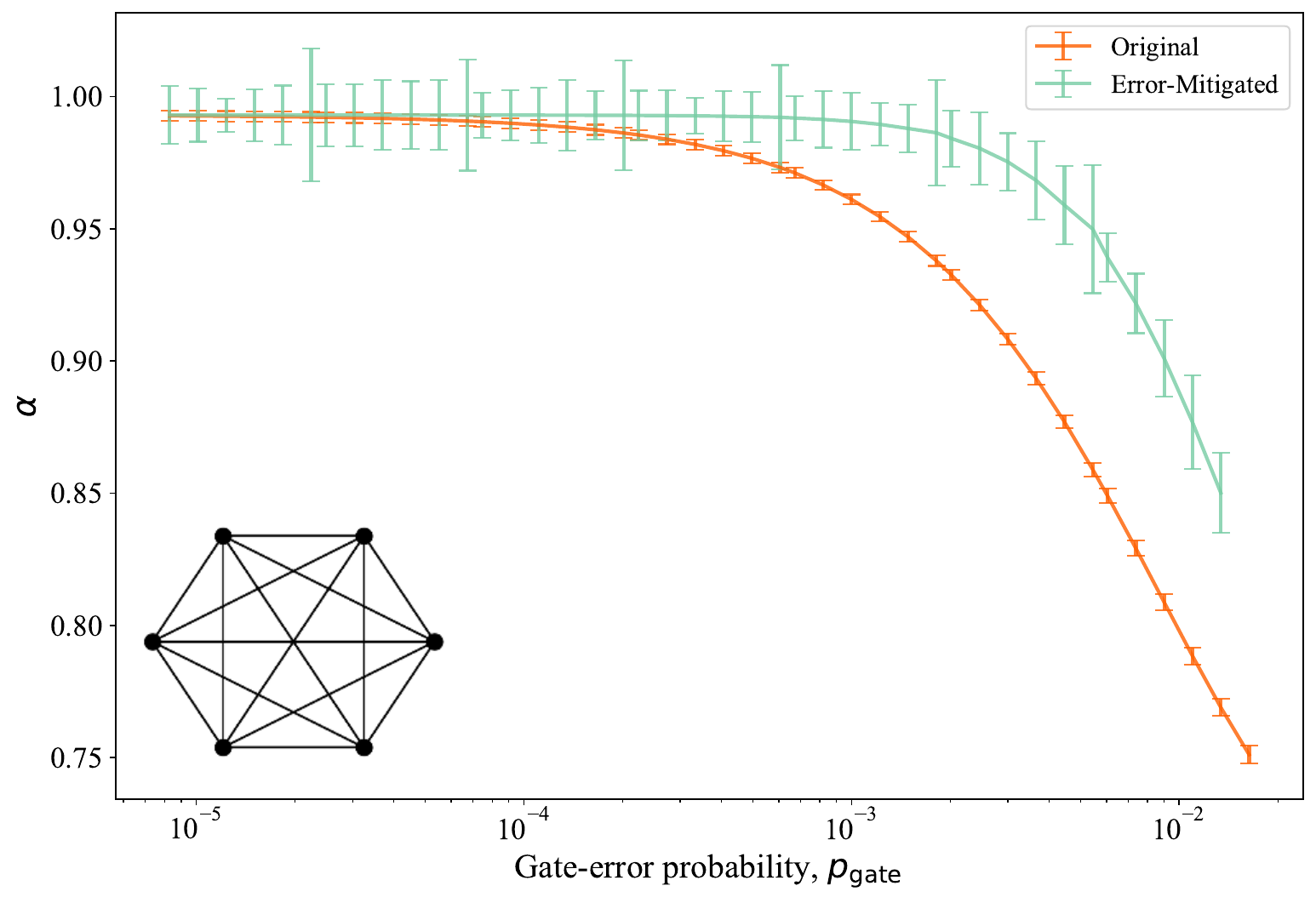}
    \caption{Final layer approximation ratio $\alpha$ as a 
    function of the gate-error probability $\pgate$, both with and 
    without error mitigation. The data were acquired 
    in noisy simulations of 6-vertex complete graphs. The error bars show the 
    standard error in the mean approximation ratio.}\label{fig:error_mitigation}
\end{figure}

In this appendix we show that error mitigation techniques can be used to improve
mean approximation ratio which can be achieved by Dynamic-ADAPT-QAOA for a 
given gate-error probability $p_\text{gate}$. In particular, we apply Richardson 
extrapolation as described in Ref~\cite{error_mitigation_2}. For a specific
value of $p_\text{gate}$ we consider the approximation ratio outputted at the end 
of the algorithm when applied to 6-vertex complete graphs. This is averaged over 100 randomized graph instances
to produce a mean approximation ratio.\\
\indent Once we have approximation ratio data for each graph, we consider what these values 
are at a different gate-error probability, $c\times p_\text{gate}$. From the data
for these two probabilities, we can extrapolate (according the formula in Ref~\cite{error_mitigation_2}),
and produce a hopefully improved approximation ratio. This procedure is repeated for all 
randomized graphs separately, after which we average the results to produce 
a mean error-mitigated approximation ratio.\\
\indent We produce curves of the mean approximation ratio against the gate-error
probability both with and without error mitigation. These are depicted in FIG.~\ref{fig:error_mitigation}.
The data shows that applying error mitigation, even if it is just to second-order, 
we can improve, on average, the outputted approximation ratio by Dynamic-ADAPT-QAOA. Additionally,
extending this technique to other versions of QAOA should be analogous. However, 
there is one important subtlety regarding the use of error mitigation to ``improve''
an algorithm's performance. In a practical setting, say in the context of Max-Cut, 
the important output is the final distribution of bit strings corresponding to 
partitions of the graph's vertices. The approximation ratio is simply used as measure of the algorithm's solution quality. Hence, although error mitigation, as applied 
above, can lead to improvements in the approximation ratio as a performance measure of solution quality, it does not actually alter the solution quality of the output bit strings which define the partition of vertices. In other words,
it is not clear what practical advantage such a technique could offer.

\end{document}